\magnification=\magstep1
\baselineskip=12pt
\font\msbm=msbm10
\centerline{{\bf p-ADIC AND ADELIC FREE RELATIVISTIC PARTICLE}}
\vskip.5cm

\centerline{G.S.DJORDJEVI\'C$^1$, B.
DRAGOVICH$^{2*}$\vfootnote{*}{E-mail:dragovic@phy.bg.ac.yu}
and LJ. NE\v SI\'C$^1$}

\vskip.5cm
\centerline{{\it $^1$Department of Physics, University of Ni\v s,P.O.Box 91,}}
\centerline{{\it 18000 Ni\v s, Yugoslavia}}
\centerline{{\it $^2$Institute of Physics, P.O.Box 57, 11001 Belgrade, 
Yugoslavia}}

\vskip1cm
\centerline{{\bf Abstract}}
\vskip.5cm

{\it We consider spectral problem for a free relativistic particle in p-adic
and adelic quantum mechanics. In particular, we found p-adic and adelic
eigenfunctions. Within adelic approach there exist   quantum
states  that exhibit  discrete structure of spacetime at the Planck scale.}

\vskip1cm

\noindent

{\bf 1. Introduction}

\vskip.5cm

Since 1987, there have been many interesting applications of p-adic
numbers in various parts of theoretical and mathematical physics (for a
review, see, e.g. Refs. 1-3). It is likely that the most attractive
investigations have been in the Planck scale physics and
non-archimedean structure of spacetime. One of the greatest
achievements in that direction is a formulation of p-adic quantum
mechanics$^{4,5}$ with its adelic generalization$^6$. Adelic quantum
mechanics unifies ordinary and p-adic ones, for all primes $p$.

There is an interplay of physical and mathematical reasons to
investigate possible role of p-adic numbers and adeles in physics.
Namely, all numerical results of experiments belong to the field of rational 
numbers $\msbm\hbox{Q}$, which is dense in the
field of real numbers $\msbm\hbox{R}$ and p-adic ones $\msbm\hbox{Q}_p$
($p$ is a prime number). $\msbm\hbox{R}$ and $\msbm\hbox{Q}_p$ $(p =
2,3,5,\cdots)$ exhaust all possible numbers which can be obtained by
completion of $\msbm\hbox{Q}$. The set of adeles $\msbm\hbox{A}$
enables to regard real and p-adic numbers simultaneosly. What is a
limit in application of real numbers in description of spacetime? Do
p-adic numbers play some role in physics? To answer these, and similar
questions, one has to construct and examine p-adic and adelic models of
quantum and relativistic physical systems.

Models at the Planck scale are of particular interest. In fact, if the
Planck length $l_0 = (\hbar G/c^3)^{1/2}\sim 10^{-33}$ cm is the
elementary one then any other length $x$ should be an integer multiple
of $l_0$. So, if one takes $l_0 = 1$ then one has $x = n$. Real
distance is $d_\infty(x,0) = \mid n\mid_\infty = n$, while the p-adic
one is $d_p(x,0) = \mid n\mid_p\le1$. Thus, p-adic geometry has to
emerge approaching the Planck scale physics.

So far, different p-adic and adelic quantum models have been studied
(see, e.g. Refs. 7-9).

In this letter we examine p-adic and adelic quantum properties of a
free relativistic particle. As a result of adelic approach we find some
discreteness of space and time at the Planck scale.

\vskip1cm

\noindent

{\bf 2. Some p-Adic and Adelic Mathematics}

\vskip.5cm

Since majority of physicists are still unfamiliar with p-adic numbers
and adeles we give here some basic facts about these attractive parts
of modern mathematics (for a profound knowledge, see, e.g. Refs. 2, 10-12).

Any p-adic number $x\in\msbm\hbox{Q}_p$ can be presented in the unique
way as an infinite expansion
$$
x = x_{-k}p^{-k}+x_{-k+1}p^{-k+1}+\cdots+x_0+x_1p+x_2p^2+\cdots, \quad
k\in\msbm\hbox{N}\ ,\eqno(2.1)
$$
where $x_i = 0,1,\cdots,p-1$ are digits. p-Adic norm of a term in (2.1)
is $\mid x_ip^i\mid_p = p^{-i}$. Since p-adic norm is the
non-archimedean one, {\it i.e.} $\mid u+v\mid_p\le \max\{\mid
u\mid_p,\mid v\mid_p\}$, it follows that $\mid x\mid_p = p^k$ in the
representation (2.1).

There are mainly two types of analysis on $\msbm\hbox{Q}_p$. The first
one ({\it i}) is based on the mapping $f:\msbm\hbox{Q}_p \to \msbm\hbox{Q}_p$, 
and the second one ({\it ii}) is related to map $\varphi:\msbm\hbox{Q}_p \to
\msbm\hbox{C}$, where $\msbm\hbox{C}$ is the set of complex numbers. We
use both of these analysis in p-adic generalization of physical models:
({\it i}) in classical and ({\it ii}) in quantum mechanics. Derivatives of 
$f(x)$ are defined as in the real case, but using p-adic norm instead of the
usual absolute value function. For mapping $\varphi(x)$ there is well-defined
integration with the Haar measure. In particular, we use the Gauss
integral$^2$
$$
\int_{\mid x\mid_{p}\le p^{\nu}}\chi_p(\alpha x^2+\beta x)dx =
\cases{p^\nu\Omega(p^\nu\mid\beta\mid_p),&$\mid\alpha\mid_p\le p^{-2\nu},$\cr
\lambda_p(\alpha)\mid2\alpha\mid_p^{-1/2}
\chi_p\big(-{\beta^2\over4\alpha}\big)\Omega\big(p^{-\nu}\mid{\beta\over2\alpha}\mid_p\big),&
$\mid4\alpha\mid_p>p^{-2\nu}.$\cr}\eqno(2.2)
$$
$\chi_p(u) = \exp(2\pi i\{u\}_p)$ is a
p-adic additive character, where $\{u\}_p$ denotes the fractional part
of $u\in\msbm\hbox{Q}_p$. $\lambda_p(\alpha)$ is an arithmetic complex-valued
function$^2$  with the following basic properties:
$$
\lambda_p(0) = 1, \ \lambda_p(a^2\alpha) = \lambda_p(\alpha), \
\lambda_p(\alpha)\lambda_p(\beta) =
\lambda_p(\alpha+\beta)\lambda_p(\alpha^{-1}+\beta^{-1}), \
\mid\lambda_p(\alpha)\mid_\infty = 1.\eqno(2.3)
$$
$\Omega(\mid u\mid_p)$ is the characteristic function on
$\msbm\hbox{Z}_p$, {\it i.e.}
$$
\Omega(\mid u\mid_p) = \cases{1,&$\mid u\mid_p\le1,$\cr
0,&$\mid u\mid_p>1,$\cr}\eqno(2.4)
$$
where $\msbm\hbox{Z}_p = \{x\in \msbm\hbox{Q}_p: \mid x\mid_p\le1\}$ is
the ring of p-adic integers.

An adele $a\in\msbm\hbox{A}$ is an infinite sequence 
$$
a = (a_\infty,a_2,\cdots,a_p,\cdots)\ ,\eqno(2.5)
$$
where $a_\infty\in\msbm\hbox{R}$ and $a_p\in\msbm\hbox{Q}_p$ with the
restriction that $a_p\in\msbm\hbox{Z}_p$ for all but a finite set $S$
of primes $p$. The set of all adeles $\msbm\hbox{A}$ can be written in
the form
$$
\msbm\hbox{A} = \mathop{U}\limits_{S}{\cal A}(S), \quad {\cal A}(S) =
\msbm\hbox{R}\times \prod_{p\in S}\msbm\hbox{Q}_p\times\prod_{p\not\in
S}\msbm\hbox{Z}_p\ .\eqno(2.6)
$$
$\msbm\hbox{A}$ is a topological space. It is a ring with respect to
componentwise addition and multiplication. There is a natural
generalization of analysis on $\msbm\hbox{R}$ and $\msbm\hbox{Q}_p$ 
to analysis on $\msbm\hbox{A}$.

\vskip1cm

\noindent

{\bf 3. Free Relativistic Particle in p-Adic Quantum Mechanics}

\vskip.5cm

In the Vladimirov-Volovich formulation$^4$ (see also Ref. 5)
one-dimensional p-adic quantum mechanics is a triple 
$$
\bigg(L_2(\msbm\hbox{Q}_p), W_p(z), U_p(t)\bigg)\ ,\eqno(3.1)
$$
where $L_2(\msbm\hbox{Q}_p)$ is the Hilbert space of complex-valued
functions of p-adic variables, $W_p(z)$ is a unitary representation of
the Heisenberg-Weyl group on $L_2(\msbm\hbox{Q}_p)$, and $U_p(t)$ is an
evolution operator on $L_2(\msbm\hbox{Q}_p)$.

$U_p(t)$ is an integral operator
$$
U_p(t)\psi_p(x) = \int_{\msbm\hbox{Q}_{p}}K_p(x,t;y,0)\psi_p(y)dy\ \eqno(3.2)
$$
whose kernel is given by the Feynman path integral
$$
K_p(x,t;y,0) = \int \chi_p\bigg(-{1\over h}S[q]\bigg){\cal D}q = \int\chi_p
\bigg(-{1\over h}\int^t_0L(q,\dot q)dt\bigg)\prod_tdq(t)\ ,\eqno(3.3)
$$
where $h$ is the Planck constant.
p-Adic Feynman path integral is investigated in Ref. 13, where it is
shown that for quadratic classical actions $\bar S(x,t;y,0)$ the solution
(3.3) becomes
$$
K_p(x,t;y,0) = \lambda_p\bigg(-{1\over2h}{\partial^2\bar S\over\partial
x\partial y}\bigg)\bigg\vert{1\over h}{\partial^2\bar S\over\partial x\partial
y}\bigg\vert_p^{1/2} \chi_p\bigg(-{1\over h}\bar S(x,t;y,0)\bigg)\ .\eqno(3.4)
$$

Expression (3.4) has the same form as that one in ordinary quantum
mechanics ({\it i.e.} with $L_2(\msbm \hbox{R})$). For a particular physical
system, p-adic eigenfunctions are subject of the spectral problem
$$
U_p(t)\psi_p^{(\alpha)}(x) = \chi_p(\alpha t)\psi_p^{(\alpha)}(x)\ .\eqno(3.5)
$$

The usual action for a free relativistic particle$^{14}$
$$
S = -mc^2\int^{\tau_{2}}_{\tau_{1}}d\tau\sqrt{\eta_{\mu\nu}\dot
x^\mu\dot x^\nu}\eqno(3.6)
$$
is nonlinear and so unsuitable for quantum-mechanical investigations.
However, a free relativistic particle can be treated as a system with
the constraint$^{15}$ $\eta_{\mu\nu}k^\mu k^\nu+m^2c^2 = k^2+m^2c^2 =
0$, which leads to the canonical Hamiltonian (with the Lagrange
multiplier $N$)
$$
H_c = N(k^2+m^2c^2)\ ,\eqno(3.7)
$$
and to the Lagrangian
$$
L = \dot x_\mu k^\mu-H_c = {\dot x^2\over4 N}-m^2c^2N\ ,\eqno(3.8)
$$
where $\dot x_\mu = \partial H_c/\partial k^\mu = 2k_\mu$ and $\dot x^2
= \dot x^\mu\dot x_\mu$. Instead of (3.6), the corresponding action for
quantum treatment of a free relativistic particle is
$$
S = \int^{\tau_{2}}_{\tau_{1}}d\tau\big({\dot x^2\over4
N}-m^2c^2N\big)\ .\eqno(3.9)
$$
From (3.9) it follows the classical trajectory
$$
\bar x^\mu =
{x_2^\mu-x_1^\mu\over\tau_2-\tau_1}\tau+{x_1\tau_2-x_2\tau_1\over
\tau_2-\tau_1}\eqno(3.10)
$$
and the classical action
$$
\bar S(x_2,T;x_1,0) = {(x_2-x_1)^2\over4T}-m^2c^2T\ ,\eqno(3.11)
$$
where $T = N(\tau_2-\tau_1)$.

All the above expressions from (3.7) to
(3.11) are valid in the real case and according to p-adic analysis
they have place  in the p-adic one. Note that the
classical action (3.11) can be presented in the form
$$
\eqalign{\bar S &= \bigg[-{(x_2^0-x_1^0)^2\over4T}-{m^2c^2T\over4}\bigg]+\bigg[
{(x_2^1-x_1^1)^2\over4T}-{m^2c^2T\over4}\bigg]\cr
&+\bigg[{(x_2^2-x_1^2)^2\over4T}-{m^2c^2T\over4}\bigg]+
{(x_2^3-x_1^3)^2\over4T}-{m^2c^2T\over4}\bigg] = \bar S^0+\bar S^1+\bar
S^2+\bar S^3\cr}\eqno(3.12)
$$ 
which is quadratic in $x_2^\mu$ and $x_1^\mu$ $(\mu = 0,1,2,3)$.

Due to (3.4) and (3.12), the corresponding quantum-mechanical
propagator may be written as product
$$
K_p(x_2,T;x_1,0) = \prod_{\mu=0}^3K_p^{(\mu)}(x_2^\mu,T;x_1^\mu,0)\ ,\eqno(3.13)
$$

$$
\eqalign{K_p^{(\mu)}(x_2^\mu,T;x_1^\mu,0) &=
\lambda_p((-1)^{\delta_{0}^{\mu}}4hT)\mid2hT\mid^{-{1\over 2}}_p \cr
&\times\chi_p\bigg[-{1\over h}(-1)^{\delta_{0}^{\mu}}{(x_2^\mu-x_1^\mu)^2\over4T}+{1\over
h}{m^2c^2T\over4}\bigg]\cr}\ ,\eqno(3.14)
$$
where $\delta_{0}^{\mu} = 1$ if $\mu = 0$ and $0$ otherwise.

Among all possible eigenstates which satisfy eq. (3.5), function
$\Omega(\mid x\mid_p)$, defined by (2.4), plays a central role in
p-adic and adelic quantum mechanics. Therefore, let us first  show
existence of $\Omega$-eigenfunction for the above relativistic
particle. In fact, we have now 1+3 dimensional problem and the
corresponding integral equation is 
$$
\int_{Q^{4}_{p}}K_p(x,T;y,0)\Omega(\mid y\mid_p)d^4y =\Omega(\mid
x\mid_p), \ (\alpha = 0)\ ,\eqno(3.15)
$$
where $\mid u\mid_p = \max_{0\le\mu\le3}\{\mid u^\mu\mid_p\}$ is p-adic
norm of $u\in \msbm\hbox{Q}_p^4$, and
$$
K_p(x,T;y,0) =  {\lambda_p^2(4hT)\over\mid2hT\mid_p^2}\chi_p\bigg(
-{(x-y)^2\over4hT}+{m^2c^2T\over h}\bigg)\ .\eqno(3.16)
$$
Eq. (3.15), rewritten in a more explicite form, reads
$$
\eqalign{&{\lambda_p^2(4hT)\over\mid2hT\mid_p^2}\chi_p\bigg({m^2c^2T\over
h}-{x^2\over4hT}\bigg)\int_{\msbm\hbox{Z}_{p}}\chi_p
\bigg({(y^0)^2\over4hT}-{x^0y^0\over2hT}\bigg)dy^0\cr
&\times\prod^3_{i=1}\int_{\msbm\hbox{Z}_{p}}\chi_p
\bigg(-{(y^i)^2\over4hT}+{x^iy^i\over2hT}\bigg)dy^i = \Omega(\mid
x\mid_p).\cr}\eqno(3.17)
$$

Using lower part of the Gauss integral (2.2) to calculate integrals in
(3.17) for each coordinate $y^\mu \ (\mu = 0,\cdots,3)$, we obtain
$$
\chi_p\bigg({m^2c^2T\over h}\bigg)\prod^3_{\mu=0}\Omega(\mid
x^\mu\mid_p) = \Omega(\mid x\mid_p), \ \ \mid hT\mid_p<1\ .\eqno(3.18)
$$
Since $\prod^3_{\mu=0}\Omega(\mid x^\mu\mid_p) = \Omega(\mid x\mid_p)$
is an identity, an equivalent assertion to (3.18) is
$$
\bigg\vert{m^2c^2T\over h}\bigg\vert_p\le 1, \ \ \ \mid hT\mid_p<1\ .\eqno(3.19)
$$

Applying also the upper part of (2.2) to (3.17), we have
$$
{\lambda_p^2(4hT)\over\mid2hT\mid_p^2}\chi_p\bigg({m^2c^2T\over
h}-{x^2\over4hT}\bigg)\prod^3_{\mu=0}\Omega\bigg(\bigg\vert{x^\mu\over2hT}\bigg\vert_p\bigg)
= \Omega (\mid x\mid_p), \ \mid4hT\mid_p\ge1\ ,\eqno(3.20)
$$
what is satisfied only for $p\not=2$. Namely, (3.20) becomes an equality if 
conditions
$$
\bigg\vert{m^2c^2T\over h}\bigg\vert_p\le1, \ \ \mid hT\mid_p = 1, \ \ p\not=2\ ,\eqno(3.21)
$$
take place.

Thus, we obtained  eigenstates
$$
\psi_p(x,T) = \cases{\Omega(\mid x\mid_p), \  \vert{m^2c^2T\over
h}\vert_p\le1,&$\mid hT\mid_p\le1, \
p\not=2,\qquad\qquad\qquad\qquad\qquad \ \ \ (3.22)$\cr
\Omega(\mid x\mid_2), \ \vert{m^2c^2T\over h}\vert_2\le1, &$\mid hT\mid_2<1,
\qquad\qquad\qquad\qquad\qquad\qquad\qquad \ (3.23)$\cr}
$$
which are invariant under  $U_p(t)$ transformation.

We have also $\Omega$-function in eigenstates
$$
\psi_p(x,T) = \chi_p\bigg({m^2c^2\over h}T\bigg)\Omega(p^\nu\mid
x\mid_p), \ \ \nu\in\msbm\hbox{Z}, \ \ \mid hT\mid_p<p^{-2\nu}\ .\eqno(3.24)
$$
This can be shown in the way similar to the previous case with
$\psi_p(x,T) = \Omega(\mid x\mid_p)$.

The eigenstates without $\Omega$-functions are as follows:
$$
\eqalign{\psi_p(x,T) = \chi_p\bigg({m^2c^2+k^2\over
h}T\bigg)\chi_p\bigg(-{kx\over h}\bigg)}, \eqno(3.25)
$$
where $k^2 = -k^0k^0+k^ik^i$, and $kx = -k^0x^0+k^ix^i$. Note that
$(m^2c^2+k^2)T = H_c\tau$ (see (3.7) and (3.11)).

\vskip1cm

\noindent

{\bf 4. Free Relativistic Particle in Adelic Quantum Mechanics}

\vskip.5cm

According to Ref. 6, the main ingredients of adelic quantum mechanics
are: ({\it i}) the Hilbert space $L_2(\msbm\hbox{A})$ of complex-valued
functions on the space of adeles $\msbm\hbox{A}$, ({\it ii}) a unitary
representation $W(z)$ of the Heisenberg-Weyl group on
$L_2(\msbm\hbox{A})$, and ({\it iii}) a unitary representation of the
evolution operator $U(t)$ on $L_2(\msbm\hbox{A})$. In a sense,
ingredients of quantum mechanics on adeles are some products of the
corresponding objects from ordinary and p-adic quantum mechanics.

So, the evolution operator is defined by
$$
U(t)\psi(x) = \int_{\msbm\hbox{A}}K(x,t;y,0)\psi(y)dy\ ,\eqno(4.1)
$$
where $t\in\msbm\hbox{A}$, $x,y\in\msbm\hbox{A}$, $\psi\in
L_2(\msbm\hbox{A})$, and 
$$
U(t)\psi(x) =
U_\infty(t_\infty)\psi_\infty(x_\infty)\prod_pU_p(t_p)\psi_p(x_p)\ .\eqno(4.2)
$$
The spectral problem is given by
$$
U(t)\psi^{(\alpha)}(x) = \chi(\alpha t)\psi^{(\alpha)}(x), \ \
\alpha\in \msbm\hbox{A}\ ,\eqno(4.3)
$$
where $\chi(\alpha t) = \chi_\infty(\alpha_\infty
t_\infty)\prod_p\chi_p(\alpha_p t_p)$ is the additive character on 
$\msbm\hbox{A}$.

Convergence of products and adelic consistency imply some conditions on
p-adic constituents of an adelic object. For instance, any
eigenfunction in (4.3) has the form
$$
\psi(x) = \psi_\infty(x_\infty)\prod_{p\in S}\psi_p(x_p)\prod_{p\not\in
S}\Omega(\mid x_p\mid_p)\ ,\eqno(4.4)
$$
where $S$ is a finite set of primes $p$. $\psi_\infty(x_\infty)\in
L_2(\msbm\hbox{R})$ and $\psi_p(x_p)$, $\Omega(\mid x_p\mid_p)\in
L_2(\msbm\hbox{Q}_p)$ are eigenfunctions of ordinary and p-adic
counterparts, respectively.

Adelic kernel of $U(t)$ for relativistic free particle is
$$
K(x,T;y,0) =
K_\infty(x_\infty,T_\infty;y_\infty,0)\prod_pK_p(x_p,T_p;y_p,0)\ ,\eqno(4.5)
$$
where $K_p(x_p,T_p;y_p,0)$ is given by (3.16), and
$K_\infty(x_\infty,T_\infty;y_\infty,0)$ is the real counterpart, which
has the same form as $K_p(x_p,T_p;y_p,0)$. The corresponding arithmetic
function $\lambda_\infty(\alpha)$ is defined by $\lambda_\infty(\alpha)
= (1-i\hbox{sgn}\alpha)/\sqrt2$ and satisfies the same basic properties as
$\lambda_p(\alpha)$ (see (2.3)).

In order to complete the adelic spectral theory for a free relativistic
particle let us now turn to the corresponding problem in ordinary
quantum mechanics in the form
$$
\int_{\msbm\hbox{R}}K_\infty(x,T_2;y,T_1)\psi_\infty(y,T_1)dy =
\chi_\infty(\alpha T_2)\psi_\infty(x)\ ,\eqno(4.6)
$$
where $\chi_\infty(a) = \exp(2\pi ia)$, $\psi_\infty(x,T) =
\chi_\infty(\alpha T)\psi_\infty(x)$ and 
$$
K_\infty(x,T_2;y,T_1) =
{\lambda_\infty^2\big(4h(T_2-T_1)\big)\over\mid2h(T_2-T_1)\mid^2_\infty}
\chi_\infty\bigg(
-{(x-y)^2\over4h(T_2-T_1)}+{m^2c^2(T_2-T_1)\over h}\bigg)\ .\eqno(4.7)
$$
(For simplicity, we omitted index $\infty$ for arguments in (4.7)).
Using the Gauss integral
$$
\int_{\msbm\hbox{R}}\chi_\infty(\alpha x^2+\beta x)dx =
\lambda_\infty(\alpha)\mid2\alpha\mid^{-1/2}_\infty
\chi_\infty\bigg(-{\beta^2\over4\alpha}\bigg), \ \ \alpha\not=0\ ,\eqno(4.8)
$$
we find the solution of (4.6) in the form
$$
\psi_\infty(x,T) = \chi_\infty\bigg({m^2c^2+k^2\over
h}T\bigg)\chi_\infty\bigg(-{kx\over h}\bigg)\ ,\eqno(4.9)
$$
where $k^2 = k_\mu k^\mu$, $kx = k_\mu x^\mu$.

From the above investigation it follows adelic eigenfunction for a free
relativistic particle:
$$
\psi(x,T) = \chi_\infty\bigg(
{m^2c^2+k^2_\infty\over h}T_\infty-{k_\infty x_\infty\over h}\bigg)
\prod_{p\in S}\psi_p(x_p,T_p)\prod_{p\not\in S}\Omega(\mid x_p\mid_p)\ ,\eqno(4.10)
$$
where $\psi_p(x_p,T)$ are given by(3.24) and (3.25). Note that
$x\in\msbm\hbox{A}^4$ in (4.10), {\it i.e.}
$$
x = \pmatrix{x^0\cr\vdots\cr x^3\cr} = \pmatrix{x_\infty^0,&x_2^0,&\ldots,&x_p^0,&\ldots\cr
\vdots&\vdots&&\vdots&&\cr
x^3_\infty,&x^3_2,&\ldots,&x^3_p,&\ldots\cr} =
(x_\infty,x_2,\cdots,x_p,\cdots)\ ,\eqno(4.11)
$$
and $k\in\msbm\hbox{A}^4$, $T\in\msbm\hbox{A}$. Any  adelic
wave function may be obtained as superposition of eigenfunctions (4.10) 
by summation over $S$ and integration over four-momentum $k$.

\vskip1cm

\noindent

{\bf 5. Concluding Remarks}

\vskip.5cm

We shown that a free relativistic particle may be regarded as a subject
not only of ordinary but also of p-adic and adelic quantum mechanics.
It is an exactly soluble theoretical model. In particular, we found
p-adic and adelic eigenfunctions.

In order to interpret adelic wave function for a free particle let us
consider its norm, {\it i.e.}
$$
\vert\psi(x,T)\vert_\infty^2 = \prod_{p\in S}\vert\psi_p(x_p,
T_p)\vert_\infty^2 \prod_{p\not\in S}\Omega(\mid x_p\mid_p)\ , \eqno(5.1)
$$
where we used $\vert \psi_\infty(x_\infty,T_\infty)\vert_\infty^2 = 1$
and $\Omega^2(\vert x_p\vert_p)\equiv\Omega(\vert x_p\vert_p)$. As
follows, (5.1) does not depend on real counterpart.

Comparison between theoretical predictions and experimental numerical
data may be done only on rational numbers. Hence, consider (5.1) in
points $x_\infty = x_2 = \cdots = x_p = \cdots = x\in\msbm\hbox{Q}$. Since
$$
\prod_{p\in S}\Omega(p^\nu\mid x\mid_p) = \cases{
1,&$x\in p^\nu\msbm\hbox{Z}$\cr
0,&$x\in\msbm\hbox{Q}\setminus p^\nu\msbm\hbox{Z},$\cr}\qquad
\prod_{p\not\in S}\Omega(\mid x\mid_p) = \cases{
1,&$x\in \msbm\hbox{Z}$\cr
0,&$x\in\msbm\hbox{Q}\setminus \msbm\hbox{Z},$\cr}\eqno(5.2)
$$
it means that $\vert\psi(x,T)\vert^2_\infty$ may be different from zero
only in a finite number of rational points which are not integers.
Extending standard interpretation of ordinary wave function to the
adelic one, it follows that the probability of finding the particle
in integer points is dominant.

There is a special (vacuum) state $(S = \emptyset)$ when all p-adic
states are $\Omega (\vert x_p\vert_p)$ and then particle can
exist only in integer points of space and time. In ordinary quantum
mechanics we label coordinates of space and time by real numbers 
which make continuum.
However, using adeles to
label space and time we obtain their discreteness in a natural way.
Conditions $\vert m^2c^2T/h\vert_p\leq 1$ and $\vert hT\vert_p\leq
\vert 2\vert_p$, which follow from (3.22) and (3.23), can be realized
choosing $h=c=m=1$ as a system of units. If $m$ is the Planck mass
then spacetime contains the Planck length as the elementary one.
Since $\vert T\vert_p = \vert N\tau\vert_p\leq\vert 2\vert_p$ and
$N$ is an arbitrary parameter, one can take $N = 2$ and obtain
$\vert \tau\vert_p\leq 1$ for every $p$. Thus, the invariant intervals
$\tau$ as well as spacetime coordinates $x^{\mu}$ are discrete.
Let us notice that there exist discrete subgroups of the Poincar\'e
group that transform discrete spacetime (lattice) into itself
(see, e.g. Ref. 16).
It is reasonable to expect that spacetime discreteness is more manifest 
in quantum gravity models and we have such situation in adelic approach 
to quantum cosmology$^{7,17}$.

Note that the above obtained results can be easily generalized to any
number of spacetime dimensions.

\vskip1cm

\noindent

{\bf References}

\vskip.5cm

\item{1.} L. Brekke and P. G. O. Freund, {\it Phys. Rep.} {\bf 233}, 1 (1993).
\item{2.} V. S. Vladimirov, I. V. Volovich and E. I. Zelenov, {\it
p-Adic Analysis and Mathematical Physics} (World Scientific, 1994).
\item{3.} A. Khrennikov, {\it p-Adic Valued Distributions in
Mathematical Physics} (Kluwer, 1994).
\item{4.} V. S. Vladimirov and I. V. Volovich, {\it Commun. Math.
Phys.} {\bf 123}, 659 (1989).
\item{5.} Ph. Ruelle, E. Thiran, D. Verstegen and J. Weyers, {\it J.
Math.Phys} {\bf 30}, 2854 (1989).
\item{6.} B. Dragovich, {\it Teor. Mat. Fiz.} {\bf 101}, 349 (1994);
{\it Int. J. Mod. Phys.} {\bf A10}, 2349 (1995).
\item{7.} B. Dragovich, "Adelic Wave Function of the Universe", in {\it
Proc. of the Third A. Friedmann Int. Seminar on Gravitation and
Cosmology}, St. Petersburg, 1996.
\item{8.} G. S. Djordjevi\'c and B. Dragovich, {\it Facta Univ.} {\bf
1}, 204 (1996).
\item{9.} B. Dragovich and Lj. Ne\v si\'c, {\it Facta Univ.} {\bf 1},
223 (1996).
\item{10.} W. H. Schikhof, {\it Ultrametric Calculus} (Cambridge Univ.
Press, 1984).
\item{11.} I. M. Gel'fand, M. I. Graev and I. I. Piatetskii-Shapiro,
{\it Representation Theory and Automorphic Functions} (Nauka, Moscow, 1966).
\item{12.} A. Weil, Adeles and Algebraic Groups ( Birkh$\ddot{\hbox{a}}$user, 1982).
\item{13.} G. S. Djordjevi\'c and B. Dragovich, {\it Mod. Phys. Lett.}
{\bf A12}, 1455 (1997).
\item{14.} B. Dragovich, P. H. Frampton and B. V. Urosevic, {\it Mod.
Phys. Lett.} {\bf A5}, 1521 (1990).
\item{15.} J. J. Halliwell and M. E. Ortiz, {\it Phys. Rev.} {\bf D48},
748 (1993).
\item{16.} P. A. M. Dirac,  Discrete Subgroups of the Poincar\'e Group,
in {\it Problems of Theoretical Physics - A Memorial Volume to Igor E.
Tamm} (Nauka, Moscow, 1972).
\item{17.} B. Dragovich, Adelic Quantum Cosmology, to be published in
{\it Proc. of the Fourth A. Friedmann Int. Seminar on Gravitation and
Cosmology}, St. Petersburg, 1998.

\end